\documentclass[prd,preprint,tightenlines]{revtex4}

\usepackage{amsmath,amssymb,bm,color,graphicx,multirow,mfirstuc}
\usepackage[colorlinks]{hyperref}
\hypersetup{
	citecolor=blue,
	linkcolor=blue,
	urlcolor=blue
}

\begin{document}
\title{Consistent explanation for the cosmic-ray positron excess in $p$-wave Breit-Wigner enhanced dark matter annihilation}

\author{Yu-Chen Ding$^{a,b}$}
\author{Yu-Lin Ku$^{a,b}$}
\author{Chun-Cheng Wei$^{a,b}$}
\author{Yu-Feng Zhou$^{a,b,c,d}$}

\affiliation{$^a$CAS key laboratory of theoretical Physics, Institute of Theoretical Physics, Chinese Academy of Sciences, Beijing 100190, China; 
$^b$School of  Physics, University of Chinese Academy of Sciences, Beijing 100049, China;
$^c$School of Fundamental Physics and Mathematical Sciences, Hangzhou Institute for Advanced Study, UCAS, Hangzhou 310024, China; 
$^d$International Center for Theoretical Physics Asia-Pacific, Beijing/Hangzhou, China.}
\date{\today}

\begin{abstract}

Dark matter (DM) annihilation in the galactic halo can be enhanced relative to that in the early Universe due to the Breit-Wigner enhancement, if the DM particles annihilate through a narrow resonance.
Although the $s$-wave Breit-Wigner enhancement can provide a consistent explanation for both the observed cosmic-ray (CR) positron excess and the DM thermal relic density, it is severely constrained by the observations of gamma rays from dwarf spheroidal satellite galaxies (dSphs) and the cosmic microwave background (CMB), which have relatively lower allowed DM annihilation cross section and typical DM velocities than that in the galactic halo.
Furthermore, in the $s$-wave Breit-Wigner enhancement, the case where the resonance mass is below a threshold (twice the DM mass) is ruled out due to the monotonically increasing annihilation cross section with decreasing DM velocity.
In this work, we consider Breit-Wigner enhanced $p$-wave DM annihilation.
We explore the parameter regions which can simultaneously account for the CR positron excess and DM thermal relic density without violating the constraints from dSphs gamma rays and CMB.
We show that the velocity-dependent cross section in this scenario can peak around the typical DM velocity in the galactic halo for the resonance mass both above and below the threshold.
Moreover, the highly suppressed annihilation cross section at extremely low DM velocity can evade the constraints from dSphs gamma rays and CMB easily, which results in larger allowed parameter regions than that in the $s$-wave case.
\end{abstract}

\maketitle

\newpage
\section{Introduction}

The existence of non-baryonic cold dark matter (DM) as the dominant component of matter in the Universe has been established by numerous astrophysical and cosmological observations, yet its particle nature remains a mystery, which constitutes one of the most profound questions in fundamental physics.
Weakly interacting massive particles (WIMPs), as a popular DM candidate, can naturally obtain the observed DM abundance through self-annihilation into the standard model (SM) final states during the process of thermal freeze-out.
The annihilation of DM into the SM particles in the local Universe, and specifically inside our Galaxy, can result in its indirect detection via energetic annihilation products such as high energy gamma rays, charged leptons, protons, and anti-protons, etc.

Recently, the AMS-02~\cite{Aguilar:2019owu} collaboration has updated their result of the cosmic-ray (CR) positron flux, which indicates a hardening at $\sim 10\ \rm{GeV}$ up to $\sim 300\ \rm{GeV}$.
The result of AMS-02 is consistent with the previous experiments of PAMELA~\cite{Adriani:2013uda} and Fermi-LAT~\cite{FermiLAT:2011ab} but with much higher accuracy.
The observed spectral feature is unexpected in conventional astrophysical theories where the majority of positrons are believed to originate from the collisions between primary CR nuclei and interstellar gas, hence strongly suggests the existence of additional positron sources.
While it is possible that some unaccounted astrophysical sources are behind this positron excess, such as nearby pulsar wind nebulae~\cite{Hooper:2008kg, Yuksel:2008rf, Profumo:2008ms, Hooper:2017gtd} and supernovae remnants~\cite{Blasi:2009hv, Hu:2009bc, Fujita:2009wk},
the DM explanation to this excess has also been widely explored~\cite{Bergstrom:2008gr, Cirelli:2008pk, Bergstrom:2009fa, Lin:2014vja, Jin:2013nta, Jin:2014ica}.
In the scenario of DM annihilation, the absence of an associated antiproton excess indicates that DM should dominantly annihilate into leptonic final states~\cite{Aguilar:2016kjl}.
The $e^+e^-$ channel leads to a very sharp spectral structure that cannot fit the observed positron flux.
For the $\mu^+\mu^-$ and $\tau^+\tau^-$ channels, a typical DM mass of $m_\chi\sim \mathcal{O}(1)\ \rm{TeV}$ and velocity-weighted annihilation cross section of $\langle \sigma_{\text{ann}} v_{\text{rel}} \rangle\sim \mathcal{O}(10^{-24})\ \rm{cm^3\ s^{-1}}$ are favored by the AMS-02 positron data~\cite{Cirelli:2008pk, Bergstrom:2009fa, Cholis:2008hb, Jin:2014ica}.

The minimal $s$-wave DM annihilation model with constant (velocity-independent) cross section has been stringently constrained by many observations.
For instance, a typical annihilation cross section of $\langle \sigma_{\text{ann}} v_{\text{rel}} \rangle\sim 3\times 10^{-26}\ \rm{cm^3\ s^{-1}}$ for WIMP DM candidate is required by the precisely measured DM relic density from Planck~\cite{Aghanim:2018eyx}, which is about two orders of magnitude lower than that required by the AMS-02 data.
This is usually referred to as the ``boost factor'' problem for DM annihilation explanation of CR positron excess~\cite{Cirelli:2008pk, Cholis:2008hb}, which is unlikely to be explained by the local DM density clumps~\cite{Diemand:2008in, Springel:2008by}.
Gamma rays produced from DM annihilation in regions with high DM density, especially in dwarf spheroidal satellite galaxies (dSphs), can be used to constrain DM properties~\cite{Ackermann:2015zua}.
Relevant observation from Fermi-LAT has set severe upper limits on DM annihilation cross section, which exclude the large cross section favored by the AMS-02 positron data for some DM annihilation channels~\cite{Lin:2014vja, Xiang:2017jou}.
During the epoch of recombination, the charged particles generated from DM annihilation could affect the temperature and polarization anisotropy of the cosmic microwave background (CMB)~\cite{Chen:2003gz, Padmanabhan:2005es, Slatyer:2015kla, Slatyer:2015jla, Galli:2009zc}.
The precise measurement performed by Planck has placed severe constraints on DM annihilation cross section~\cite{Aghanim:2018eyx}.
The constraints from CMB are more stringent than that from dSphs gamma rays for some DM annihilation channels, and are free of some astrophysical uncertainties compared to the results from CR observations, which arise from the large-scale structure formation, DM density profile, etc.~\cite{Slatyer:2015jla}.
To reconcile the contradictions mentioned above, a more detailed DM annihilation model is required.

The typical velocity of DM particles in our galactic halo ($\sim 10^{-3}c$, with $c$ the speed of light) is much higher than that in nearby dSphs ($\sim 10^{-4}c$) and during recombination ($\sim 10^{-8}c$), but two orders of magnitude lower than that during thermal freeze-out of WIMPS.
Therefore, if the DM annihilation cross section is velocity-dependent in a non-monotonic way and peaks around the typical DM velocity in the galactic halo, it can survive the constraints from the CMB and dSphs gamma rays, and still generate the desired boost factor.
In our recent work~\cite{Ding:2021zzg}, we considered the Sommerfeld mechanism to reconcile the contradictions mentioned above, which occurs when the annihilating particles at low velocity self-interact via a long-range attractive potential~\cite{https://doi.org/10.1002/andp.19314030302,Hisano:2003ec,Hisano:2002fk,Cirelli:2007xd,ArkaniHamed:2008qn,Pospelov:2008jd,March-Russell:2008klu,Iengo:2009ni,Cassel:2009wt}.
Although the $s$-wave Sommerfeld-enhanced DM annihilation is highly constrained due to the monotonically increasing cross section with decreasing DM velocity,
the $p$-wave Sommerfeld-enhanced DM annihilation is still a viable possibility.
Due to the additional velocity-square factor in $p$-wave annihilation cross section and the saturation of Sommerfeld enhancement under a certain DM velocity, 
the underlying DM annihilation cross section can naturally have the desired peak structure.

Breit-Wigner enhancement is another extensively studied mechanism for velocity-dependent cross section, which occurs when particles annihilate via a narrow resonance~\cite{Ibe:2008ye,Guo:2009aj,Bi:2009uj,Bi:2011qm,Bai:2017fav,Li:2015tka,Xiang:2017jou}.
In Ref.~\cite{Xiang:2017jou}, Breit-Wigner enhanced $s$-wave DM annihilation was considered to reconcile the contradictions mentioned above.
The case where the resonance mass is below a threshold (twice the DM mass) is ruled out due to the monotonically increasing cross section with decreasing DM velocity.
When the resonance mass is above the threshold, parameter regions exist to account for these observations consistently but only the parameters with extremely tiny values of mass deviation from the threshold and decay width of the resonance can survive.

In this work, we consider DM particles annihilating through a $p$-wave process with Breit-Wigner enhancement.
We take into account the impact of the kinetic decoupling effect on DM thermal relic density.
We systematically calculate the velocity-dependent astrophysics factor ($J$-factor) for the $p$-wave Breit-Wigner enhanced DM annihilation for 15 nearby dSphs and derive the corresponding upper limits of the DM annihilation cross section with the dSphs gamma-ray data from Fermi-LAT.
Through a combined analysis of the observation data of CR positron excess, DM thermal relic density, dSphs gamma rays, and CMB, we obtain the parameter regions that can explain all these observables consistently in the $p$-wave Breit-Wigner enhanced DM annihilation model.
We find that, unlike the $s$-wave scenario where the parameter regions that can explain the large boost factor are severely limited by the constraints from dSphs gamma rays and CMB, in the $p$-wave scenario, due to the extra velocity-square factor, the cross section is highly suppressed at extremely low DM velocity hence the constraints from dSphs gamma rays and CMB are alleviated.
In addition, whether the resonance mass is above or below the threshold, there exist parameter regions that can account for all these observations simultaneously.
For the same reason as in our recent work~\cite{Ding:2021zzg}, we do not consider the constraints from other observables, such as the gamma rays from the galactic center (GC) observed by H.E.S.S.~\cite{Abdallah:2016ygi} and the isotropic gamma ray background (IGRB) measured by Fermi-LAT~\cite{Ackermann:2015tah} due to the additional uncertainties they involve.

This paper is organized as follows.
In Sec.~\ref{sec:ams}, we derive the AMS-02 favored DM mass and annihilation cross section through fitting to the updated AMS-02 CR positron data in the DM annihilation scenario.
In Sec.~\ref{sec:bw}, we briefly describe the Breit-Wigner enhancement mechanism.
In Sec.~\ref{sec:relic}, we discuss the constraints from the observed DM thermal relic density.
In Sec.~\ref{sec:fermi-lat}, we calculate the upper limits of the DM annihilation cross section from gamma-ray data of 15 nearby dSphs measured by Fermi-LAT.
In Sec.~\ref{sec:cmb}, we discuss the CMB bound on the DM annihilation cross section given by Planck.
In Sec.~\ref{sec:result}, we identify the parameter regions that can simultaneously explain all these relevant observations.
We summarize our conclusions in Sec.~\ref{sec:conclusion}.

\section{DM annihilation explanation of the AMS-02 positron excess}\label{sec:ams}

The propagation of CR positrons in the galactic halo can be described by diffusion models~\cite{Ginzburg:1990sk,Strong:2007nh}.
For the CR source and propagation, we adopt the same model from our previous work in Ref.~\cite{Ding:2021zzg}.
We use the public code \texttt{GALPROP v54}~\cite{Strong:1998pw,Moskalenko:2001ya,Strong:2001fu, Moskalenko:2002yx,Ptuskin:2005ax} to solve the propagation equation numerically.
We focus on the annihilation processes of DM particles $\chi$ to final states of two leptons $\mu^+\mu^-$ and $\tau^+\tau^-$.
The Monte-Carlo event generator \texttt{PYTHIA 8.2}~\cite{Sjostrand:2014zea} is used to simulate the positron spectrum produced by DM annihilation.
The DM particle mass $m_\chi$, velocity-weighted annihilation cross section $\langle \sigma_\text{ann} v_\text{rel}\rangle$, and an extra normalization factor of the background flux of CR positrons $C_{e^+}$ are free parameters
to be determined.
We use the Bayesian analysis to perform a global fit to the latest AMS-02 positron data~\cite{Aguilar:2019owu}.
The prior distributions of the free parameters are uniform distributions whose prior ranges are listed in Tab.~\ref{tab:best-fit}. 
The likelihood function is assumed to be Gaussian
\begin{equation}
\mathcal{L} = \prod_i \rm{exp}[-\frac{(\Phi_i-\Phi_{\rm{exp,i}})^2}{2\sigma^2_{\rm{exp,i}}}]\ ,
\end{equation}
where $\Phi_i$ is the predicted value, $\Phi_{\rm{exp,i}}$ and $\sigma_{\rm{exp,i}}$ are the measured value and uncertainty of the CR positron flux in energy bin $i$.
To efficiently explore the parameter space, we adopt the \texttt{MultiNest} sampling algorithm~\cite{ Feroz:2007kg, Feroz:2008xx, Feroz:2013hea}.
We use $\chi^2=-2\rm{ln}\mathcal{L}$ to characterize the goodness-of-fit.

\begin{table}[t]
\begin{tabular}{c|cc|cc|cc|c}
\hline \hline
\multirow{3}{*}{channel} & \multicolumn{2}{c|}{$\rm{log_{10}}(m_\chi/\rm{GeV})$} & \multicolumn{2}{c|}{$\rm{log_{10}}(\langle\sigma_{\text{ann}} v_{\text{rel}}\rangle/\rm{cm^3\ s^{-1}})$} & \multicolumn{2}{c|}{$C_{e^+}$} & \multirow{3}{*}{$\chi^2/\text{d.o.f}$} \\
\cline{2-7} & \multicolumn{2}{c|}{Prior range: [1, 4]} & \multicolumn{2}{c|}{Prior range: [-26, -21]} & \multicolumn{2}{c|}{Prior range: [0.1, 10]} \\
\cline{2-7} & (Mean, $\sigma$) & Best-fit & (Mean, $\sigma$) & Best-fit & (Mean, $\sigma$) & Best-fit \\
\hline
$\mu^+\mu^-$ & $2.85 \pm 0.02$ & 2.85 & $-23.65 \pm 0.03$ & -23.66 & $1.69 \pm 0.02$ & 1.69 & 52.22/32 \\
$\tau^+\tau^-$ & $3.24 \pm 0.04$ & 3.24 & $-22.81 \pm 0.05$ & -22.8 & $1.53 \pm 0.04$ & 1.53 & 21.46/32 \\
\hline \hline
\end{tabular}
\caption{Prior ranges, posterior means, standard deviations, and best-fit values of $m_\chi$, $\langle\sigma_{\text{ann}} v_{\text{rel}}\rangle$, and $C_{e^+}$ for $\mu^+\mu^-$ and $\tau^+\tau^-$ channels. The values of $\chi^2/\text{d.o.f}$ are also listed as an estimation of the goodness-of-fit.}
\label{tab:best-fit}
\end{table}

\begin{figure}[t]
\includegraphics[width=0.505\textwidth]{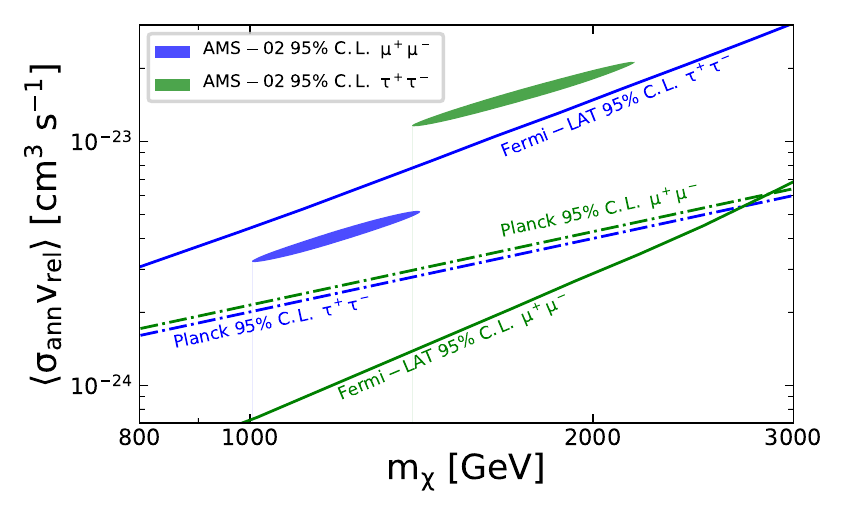}
\includegraphics[width=0.475\textwidth]{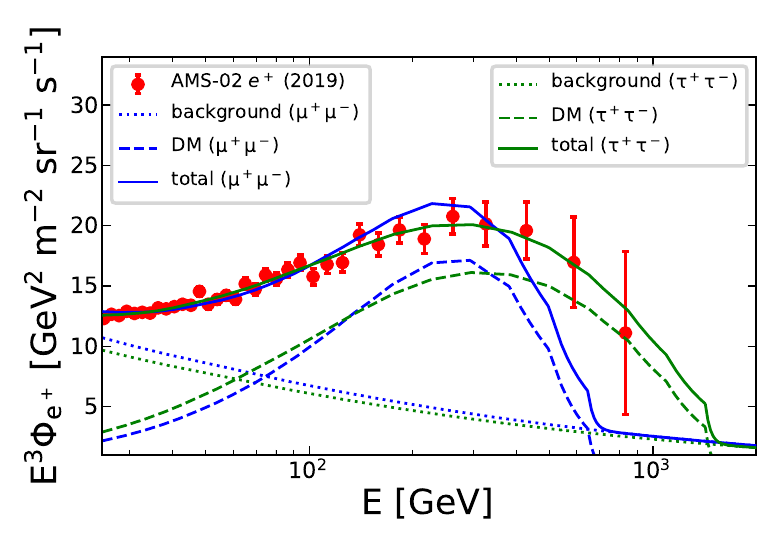}
\caption{
Left) Regions of DM mass and annihilation cross section favored by the AMS-02 positron data~\cite{Aguilar:2019owu} at 95\%~C.L.
The blue (green) region corresponds to the annihilation channel of $\mu^+\mu^-$ ($\tau^+\tau^-$).
In the assumption of velocity-independent DM annihilation cross section,
the solid blue (green) curve represents the upper limits at $95\%$ C.L. derived from the Fermi-LAT data on dSphs gamma rays for the $\mu^+\mu^-$ ($\tau^+\tau^-$) channel, which are taken from Ref.~\cite{Ackermann:2015zua},
and the dashed blue (green) curve stands for the upper limit at $95\%$ C.L. derived from the Planck measurements of CMB~\cite{Aghanim:2018eyx} for the $\mu^+\mu^-$ ($\tau^+\tau^-$) channel.
Right) Best-fit CR positron fluxes (solid lines) for two different annihilation channels. The dashed and dotted blue curves stand for the contributions from DM annihilation and secondary positron backgrounds. The blue and green curves correspond to the $\mu^+\mu^-$ and $\tau^+\tau^-$ channels, respectively.
}
\label{fig:ams-pos}
\end{figure}

The best-fit values, posterior means, and standard deviations of the fitting parameters and the goodness-of-fit for $\mu^+\mu^-$ and $\tau^+\tau^-$ channels are summarized in Tab.~\ref{tab:best-fit}.
In the left panel of Fig.~\ref{fig:ams-pos}, we show the regions favored by the AMS-02 positron data in the $(m_\chi, \langle\sigma_{\text{ann}} v_{\text{rel}}\rangle)$ plane at $95\%$ C.L., which suggest that $m_\chi\approx 1-1.4\ \rm{TeV}$ with $\langle\sigma_{\text{ann}} v_{\text{rel}}\rangle\approx 3.2-5.2\times 10^{-24}\ \rm{cm^3\ s^{-1}}$ is favored for the $\mu^+\mu^-$ channel and $m_\chi\approx 1.4-2.2\ \rm{TeV}$ with $\langle\sigma_{\text{ann}} v_{\text{rel}}\rangle\approx 1.1-2.1\times 10^{-23}\ \rm{cm^3\ s^{-1}}$ for the $\tau^+\tau^-$ channel.
The upper limits at $95\%$ C.L. derived from the Fermi-LAT data on dSphs gamma rays~\cite{ Ackermann:2015zua} and the Planck measurement of CMB~\cite{Aghanim:2018eyx} (see Sec.~\ref{sec:cmb} for details) in the assumption of constant DM annihilation cross section are also shown for comparison.
The AMS-02 favored regions are compatible with the constraints from dSphs gamma rays in $\mu^+\mu^-$ channel, but in strong tension with that in $\tau^+\tau^-$ channel due to tremendous photons from the hadronic decay of tauons.
In both channels, the AMS-02 favored regions are totally excluded by the constraints from CMB.
In the right panel of Fig.~\ref{fig:ams-pos}, we show the calculated CR positron flux with the best-fit DM parameters, as well as the secondary positron backgrounds.
The positron spectra from DM annihilation through the $\mu^+\mu^-$ channel are relatively narrow compared with the observations.
Nevertheless, DM annihilation through the $\tau^+\tau^-$ channel predicts broader spectra, which is in good agreement with the AMS-02 positron data.
This can also be seen from the values of $\chi^2/\text{d.o.f}$ in Tab.~\ref{tab:best-fit}.
Hence, we mainly focus on the $\tau^+\tau^-$ channel in the following analysis.
The resonance coupling dominantly to tauons can be achieved via the flavor-specific scenarios which allow the mediator to couple dominantly to one flavor in a technically natural way~\cite{Batell:2017kty, Marsicano:2018vin}, or the lepton-specific scenarios in which the coupling of the mediator to one flavor is proportional to the corresponding lepton mass~\cite{Batell:2016ove, DAmbrosio:2002vsn}.

\section{Breit-Wigner enhancement}\label{sec:bw}

We consider a simplified model where two fermionic DM particles directly annihilate into a pair of charged leptons via a narrow resonance $R$.
We adopt two auxiliary parameters $\delta$ and $\gamma$ to express the mass $M$ and decay width $\Gamma$ of the resonance
\begin{equation}
\delta = 1-M^2/4m^2_\chi\ \rm{and}\ \gamma=\Gamma/M\ .
\end{equation}
When the resonance mass is close to twice the DM mass and the decay width is narrow, the annihilation cross section is sensitive to the relative velocity of incident DM particles $v_{\rm{rel}}$.
For a vector resonance, the DM annihilation is an $s$-wave process and the corresponding cross section has a general formula of~\cite{Ibe:2008ye, Bi:2009uj, Bi:2011qm, Xiang:2017jou}
\begin{equation}
\sigma_{\rm{ann}} v_{\rm{rel}} = \frac{a}{(\delta +v_{\rm{rel}}^2/4c^2)^2+\gamma ^2}\ ,
\end{equation}
where $a$ is a global factor.
For a scalar resonance, the DM annihilation is a $p$-wave process and the corresponding cross section has an extra velocity-square factor.
For simplicity, we parameterize the cross section with a global factor $b$ as
\begin{equation} \label{cs}
\sigma_{\rm{ann}} v_{\rm{rel}} = \frac{b v_{\rm{rel}}^2/c^2}{(\delta +v_{\rm{rel}}^2/4c^2)^2+\gamma ^2}\ .
\end{equation}
In both cases, for a given relative velocity of incident DM particles, when $\delta>0$, namely, $M<2m_\chi$, the DM annihilation cross section increases monotonically with decreasing $\delta$ and $\gamma$.
However, for the case of $\delta<0$, namely, $M>2m_\chi$,
the DM annihilation cross section remains increasing monotonically with decreasing $\gamma$ while reaching a maximum at $\delta=-v_{\rm{rel}}^2/4c^2$.
In the case of vector resonance, the annihilation cross section scales as $c^4/v^4_{\rm{rel}}$ when $v_{\rm{rel}}^2/c^2\gg |\delta|,\gamma$, and remains almost unchanged with $v_{\rm{rel}}$ due to the saturation of the Breit-Wigner enhancement when $v_{\rm{rel}}^2/c^2\ll |\delta|,\gamma$.
In the case of scalar resonance,
when $v_{\rm{rel}}^2/c^2\gg |\delta|,\gamma$, the annihilation cross section scales as $c^2/v^2_{\rm{rel}}$.
When $v_{\rm{rel}}^2/c^2\ll |\delta|,\gamma$, the Breit-Wigner enhancement saturates, and the annihilation cross section scales as $v^2_{\rm{rel}}/c^2$.

\section{Constraints from DM thermal relic density}\label{sec:relic}

\begin{figure}[t]
\includegraphics[width=0.7\textwidth]{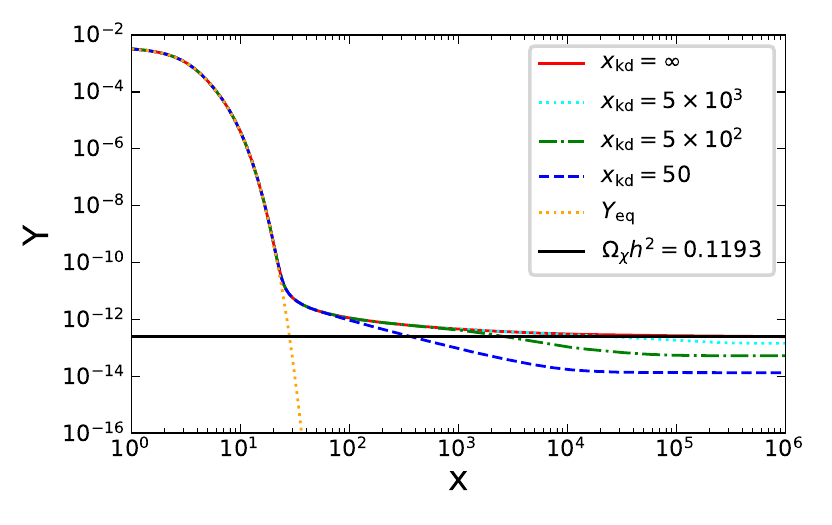}
\caption{The evolution of DM density $Y$ as a function of $x$.
The blue, green, cyan, and red lines correspond to $x_{\rm{kd}}=50,\ 5\times 10^2,\ 5\times 10^3,\ \infty$, respectively. Parameters are set at $\delta = 10^{-7},\ \gamma = 10^{-7}$, $m_\chi=1.7\ \rm{TeV}$, and $b$ is fixed to reproduce the correct DM thermal relic density in the case without kinetic decoupling. The orange line is the evolution of $Y_{\rm{eq}}$, and the black line is the observed DM relic abundance from Planck~\cite{Aghanim:2018eyx}.
}  
\label{fig:Yx}
\end{figure}

\begin{figure}[t]
\includegraphics[width=\textwidth]{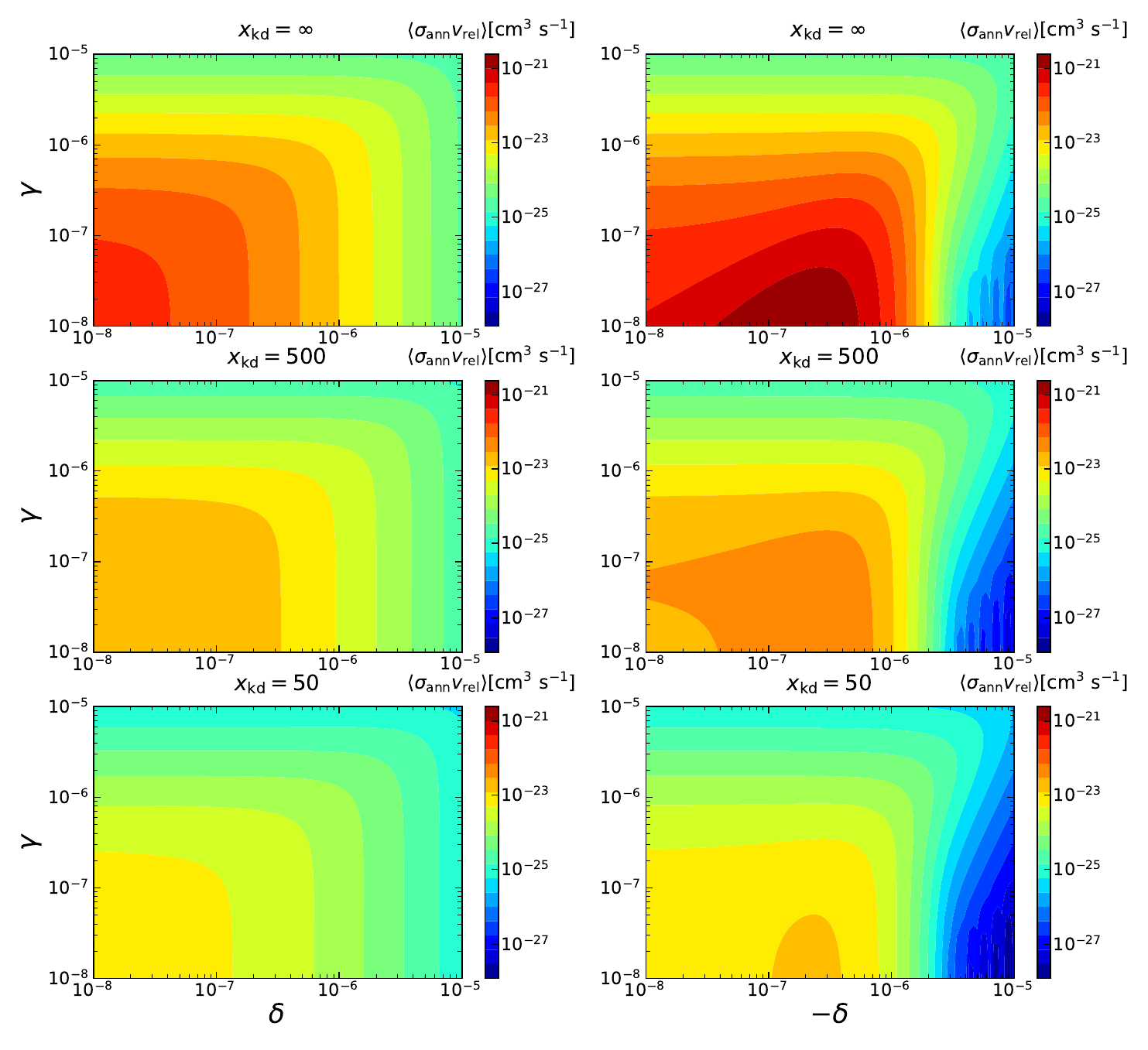}
\caption{
Value of $\left<\sigma_{\rm{ann}} v_{\rm{rel}}\right>$ calculated from Eq.~(\ref{eq:cs}) with $v_0=220\ \rm{km\ s^{-1}}$ as a function of $\delta$ and $\gamma$, where DM mass is fixed at the best-fit value of AMS-02 positron data in $\tau^+\tau^-$ channel $m_\chi=1.7\ \rm{TeV}$ and $b$ is derived from the correct DM thermal relic density. $x_{\rm{kd}}$ in the lower, middle, and upper two panels are $50,\ 500,\ \infty$, respectively. The left column is for the case of $\delta>0$, and the right column is for the case of $\delta<0$.
}
\label{fig:relic}
\end{figure}

The thermal evolution of DM density can be described by the Boltzmann equation~\cite{Kolb:1988aj}
\begin{equation}
\label{eq:Bolzmann}
\frac{d Y}{d x}= - \sqrt{\frac{\pi}{45}}m_{\rm{pl}} m_{\chi}g_{\ast}^{-1/2} g_{\ast s} 
x^{-2}\left<\sigma_{\rm{ann}} v_{\rm{rel}}\right>\left(Y^2-Y_{eq}^2\right)\ ,
\end{equation}
where $Y = n_\chi /s$ is the DM number density $n_\chi$ normalized by the entropy density $s$, $Y_{\rm{eq}}=0.145(g/g_{\ast s})x^{3/2}e^{-x}$ is the DM density in equilibrium~\cite{Jungman:1995df}, $g$ is the number of degrees of freedom of the DM particle, $x = m_\chi/T$, $T$ is the temperature of the thermal bath, $m_{\rm{pl}} \approx 1.22\times 10^{19}\ \rm{GeV}$ is the Planck mass, $g_{\ast}$ and $g_{\ast s}$ are the effective relativistic degrees of freedom for energy and entropy density, respectively~\cite{Steigman:2012nb}.
If the DM particles are in thermal equilibrium with temperature $T_\chi$, the velocity distribution of DM particles is the Maxwell-Boltzmann distribution, characterized by the most probable velocity $v_0$, which is related to the DM temperature as $v_0/c=\sqrt{2T_\chi/m_\chi}$.
Then, the velocity-averaged cross section of the $p$-wave Breit-Wigner enhanced DM annihilation can be expressed as
\begin{equation} 
\langle \sigma_{\rm{ann}} v_{\rm{rel}}\rangle =\frac{b}{v_0^3} \sqrt{\frac{2}{\pi}} \int_0^{+\infty} \mathrm{d}v_{\rm{rel}}\ v_{\rm{rel}}^2 e^{-\frac{v_{\rm{rel}}^2}{2v_0^2}} \frac{v_{\rm{rel}}^2/c^2}{(\delta+v_{\rm{rel}}^2/4c^2)^2+\gamma^2}\ .
\label{eq:cs}
\end{equation}
The DM relic abundance has a relation with the value of $Y$ in the present day ($x=x_\infty$) as~\cite{Chen:2013bi}
\begin{equation}
\Omega_{\chi} h^2 \approx 2.76 \times 10^8 Y(x_\infty)\left(\frac{m_\chi}{\rm{GeV}}\right)\ ,
\end{equation}
and the corresponding observed value from Planck is $\Omega_{\chi}h^2\simeq0.1193$~\cite{Aghanim:2018eyx}.

After freeze-out (when $Y-Y_{\rm{eq}}=\mathcal{O}(Y_{\rm{eq}})$, typically, $x_f\equiv m_\chi/T_f\sim \mathcal{O}(10)$), DM particles remain kinematically coupled to the thermal bath through elastic scattering. 
Kinetic decoupling occurs later, when the momentum transfer rate drops below the Hubble expansion rate.
The temperature of DM kinetic decoupling $T_{\rm{kd}}$ depends on the model details of DM annihilation~\cite{Bi:2011qm, Bai:2017fav}.
To perform a model-independent analysis, we take $T_{\rm{kd}}$ as a free parameter in this work.
Before the kinetic decoupling of DM, the DM particles have the same temperature with the thermal bath $T_\chi=T$.
After that, $T_\chi$ drops as $a^{-2}$, while $T$ drops as $a^{-1}$, where $a$ is the scale factor of the Universe, and so $T_\chi=T^2/T_{\rm{kd}}$~\cite{Chen:2001jz, Hofmann:2001bi, Bringmann:2006mu, Bi:2011qm}.

In Fig.~\ref{fig:Yx}, we show the evolution of DM density with different values of DM kinetic decoupling temperature $x_{\rm{kd}}\equiv m_\chi/T_{\rm{kd}}=50,\ 5\times 10^2$ and $5\times 10^3$.
We also give the results in the limit of $x_{\rm{kd}}=\infty$, corresponding to the case without kinetic decoupling.
After kinetic decoupling, the DM annihilation becomes more significant due to the more rapid decrease in the DM temperature and more significant increase in Breit-Wigner enhancement, which reduces the DM abundance more efficiently.
If the kinetic decoupling occurs much later than the freeze-out, namely, $x_{\rm{kd}}\gg x_f$, such effect becomes not important compared with the case without kinetic decoupling.
However, if the kinetic decoupling occurs at nearly the same epoch as that for the freeze-out, the efficient DM annihilation would reduce DM relic abundance by about one order of magnitude.

Under the requirement of reproducing the observed DM relic density, the value of $b$ in the expression of the DM annihilation cross section can be determined for given parameters $m_\chi$, $\delta$ and $\gamma$.
For a given velocity distribution of DM particles characterized by the most probable velocity $v_0$, the $p$-wave Breit-Wigner enhanced DM annihilation cross section $\left<\sigma_{\rm{ann}} v_{\rm{rel}}\right>$ can be calculated from the determined value of $b$ through Eq.~(\ref{eq:cs}).
In Fig.~\ref{fig:relic}, we show the predicted value of $\left<\sigma_{\rm{ann}} v_{\rm{rel}}\right>_{\rm{halo}}$ with $v_0=v_{\rm{halo}}\approx 220\ \rm{km\ s^{-1}}$ in the galactic halo as a function of $\delta$ and $\gamma$  after the constraints from the DM thermal relic density, for three typical values of DM kinetic decoupling temperature $x_{\rm{kd}}=50,\ 500,\ \infty$ in both cases of $\delta>0$ and $\delta<0$.
The DM mass is fixed at the best-fit value $m_\chi=1.7\ \rm{TeV}$ for AMS-02 positron data in the $\tau^+\tau^-$ channel.
In the case of $\delta>0$, although smaller values of $\delta$ and $\gamma$ lead to stronger Breit-Wigner enhancement at freeze-out and hence a smaller $b$, corresponding predicted value of $\left<\sigma_{\rm{ann}} v_{\rm{rel}}\right>_{\rm{halo}}$ is larger due to the rapid increase of the rescale factor in Eq.~(\ref{eq:cs}) with decreasing $\delta$ and $\gamma$.
In the case of $\delta<0$, $\left<\sigma_{\rm{ann}} v_{\rm{rel}}\right>_{\rm{halo}}$ reaches a maximum at $\delta\sim -v_{\rm{halo}}^2/c^2$ and $\gamma\lesssim v_{\rm{halo}}^2/c^2$, which is also derived from the behavior of the rescale factor in Eq.~(\ref{eq:cs}).
The figure shows that the case of $\delta<0$ allows larger boost factor than the case of $\delta>0$.
As mentioned earlier, since a smaller $x_{\rm{kd}}$ (or higher kinetic decoupling temperature) reduces the DM relic abundance more efficiently,
corresponding value of $b$ and hence the value of $\left<\sigma_{\rm{ann}} v_{\rm{rel}}\right>_{\rm{halo}}$ are required to be smaller in order to reproduce the observed DM thermal relic density.

\section{Constraints from dSphs gamma-ray observations from Fermi-LAT}
\label{sec:fermi-lat}

\subsection{Gamma-ray flux from DM annihilation in dSphs and $p$-wave Breit-Wigner enhanced $J$-factor}

The dSphs of the Milky Way are promising targets for the indirect detection of DM annihilation due to their large DM concentration, low diffuse galactic gamma-ray foreground, and lack of conventional astrophysical gamma-ray production mechanisms~\cite{Mateo:1998wg, Grcevich:2009gt}.
The gamma-ray flux from DM self-annihilation within an energy range ($E_{\min}$, $E_{\max}$) from a solid angle $\Delta\Omega$ can be expressed as~\cite{Ding:2021zzg}
\begin{equation}\label{eq:gamma_flux} 
	\Phi_\gamma(\Delta \Omega, E_{\min}, E_{\max}) =
 	\frac{C}{8\pi m_{\chi}^{2}} 
 	\int_{E_{\min}}^{E_{\max}} 
 	\frac{dN_\gamma}{dE_\gamma}{dE_\gamma}\times J\ ,
\end{equation}
where $C$ is the velocity-independent part of the DM annihilation cross section, and the $J$-factor contains the information of the DM distribution and the velocity-dependent part of the cross section.
In the simplest case where the DM annihilation cross section is velocity independent, $C=\sigma_{\rm{ann}} v_{\rm{rel}}$, and the corresponding $J$-factor is given by
\begin{equation} \label{eq:J0}
J_0=\int_{\Delta \Omega} d \Omega \int_{\text{l.o.s}}  \rho^{2}(\mathbf{r}) dl\ ,
\end{equation}
where $\rho(\mathbf{r})$ is the DM density distribution in dSphs.
The $J_0$ factor contains only the astrophysical information. 
In the case of $s$-wave Breit-Wigner enhanced DM annihilation, $C=a$, and the corresponding $s$-wave $J$-factor can be written as
\begin{equation} \label{eq:Js}
 		J_s=\int_{\Delta \Omega} d \Omega \int_{l.o.s} dl \int d^{3} \boldsymbol{v}_1 \int d^{3} \boldsymbol{v}_2 f\left(\boldsymbol{r}, \boldsymbol{v}_1\right) f\left(\boldsymbol{r}, \boldsymbol{v}_2\right) \frac{1} {(\delta+v_{\text{rel}}^2/4c^2)^2+\gamma^2}\ ,
\end{equation}
where $f\left(\boldsymbol{r}, \boldsymbol{v}\right)$ is the phase-space distribution of DM particles within dSphs.
Similarly, in the case of $p$-wave Breit-Wigner enhanced DM annihilation, $C=b$, and the corresponding $p$-wave $J$-factor is
\begin{equation} \label{eq:Jp}
 		J_p=\int_{\Delta \Omega} d \Omega \int_{l.o.s} dl \int d^{3} \boldsymbol{v}_1 \int d^{3} \boldsymbol{v}_2 f\left(\boldsymbol{r}, \boldsymbol{v}_1\right) f\left(\boldsymbol{r}, \boldsymbol{v}_2\right) \frac{v_{\text{rel}}^2/c^2} {(\delta+v_{\text{rel}}^2/4c^2)^2+\gamma^2}\ .
 \end{equation}

 \begin{table}
\begin{tabular}{ c c c c c c }
\hline \hline
\multirow{2}{*}{Name} & $J_p\ [\times 10^{25}]$ & $J_p\ [\times 10^{25}]$ & $J_p\ [\times 10^{21}]$ & $J_p\ [\times 10^{17}]$ & \multirow{2}{*}{$J_0\ [\times 10^{18}]$} \\
& $(\delta, \gamma) = (-10^{-8}, 10^{-8})$ & $(10^{-8}, 10^{-8})$ & $(10^{-6}, 10^{-6})$ & $(10^{-4}, 10^{-4})$ &  \\
\hline
Bootes I 		  & $19.0$ & $11.9$ & $14.9$ & $15.0$ & $6.3$ \\
Canes Venatici II & $1.3$ & $1.0$ & $1.1$ & $1.1$ & $0.8$ \\
Carina 			  & $2.7$ & $2.0$ & $2.3$ & $2.3$ & $1.3$ \\
Coma Berenices    & $17.0$ & $12.3$ & $14.5$ & $14.6$ & $10.0$\\
Draco 			  & $36.6$ & $19.1$ & $26.1$ & $26.3$ & $6.3$ \\
Fornax 			  & $8.7$ & $4.3$ & $6.1$ & $6.3$ & $1.6$ \\
Hercules		  & $2.5$ & $1.7$ & $2.1$ & $2.1$ & $1.3$ \\
Leo II 			  & $0.9$ & $0.6$ & $0.8$ & $0.8$ & $0.4$ \\
Leo IV 			  & $1.9$ & $1.3$ & $1.5$ & $1.5$ & $0.8$ \\
Sculptor		  & $18.8$ & $10.5$ & $13.9$ & $13.7$ & $4.0$ \\
Segue 1 		  & $54.7$ & $41.6$ & $46.9$ & $47.0$ & $31.6$ \\
Sextans 		  & $7.7$ & $5.0$ & $6.3$ & $6.3$ & $2.5$ \\
Ursa Major II 	  & $46.8$ & $32.3$ & $39.8$ & $39.8$ & $20.0$ \\
Ursa Minor 		  & $39.9$ & $19.5$ & $28.2$ & $28.3$ & $6.3$ \\
Willman 1 		  & $21.8$ & $16.4$ & $18.9$ & $18.7$ & $12.6$ \\
\hline \hline
\end{tabular}
\caption{The values of $p$-wave Breit-Wigner enhanced $J$-factor $J_p$ (in units of $\mathrm{GeV^2\ cm^{-5}}$) of 15 dSphs at four typical values of $(\delta, \gamma)=(-10^{-8}, 10^{-8}), (10^{-8}, 10^{-8}), (10^{-6}, 10^{-6})$, and $(10^{-4}, 10^{-4})$. The last column is the values of the normal $J$-factor $J_0$ taken from~\cite{ Ackermann:2015zua}.
}
\label{tab:js}
\end{table}

\begin{figure}[t]
\includegraphics[width=\textwidth]{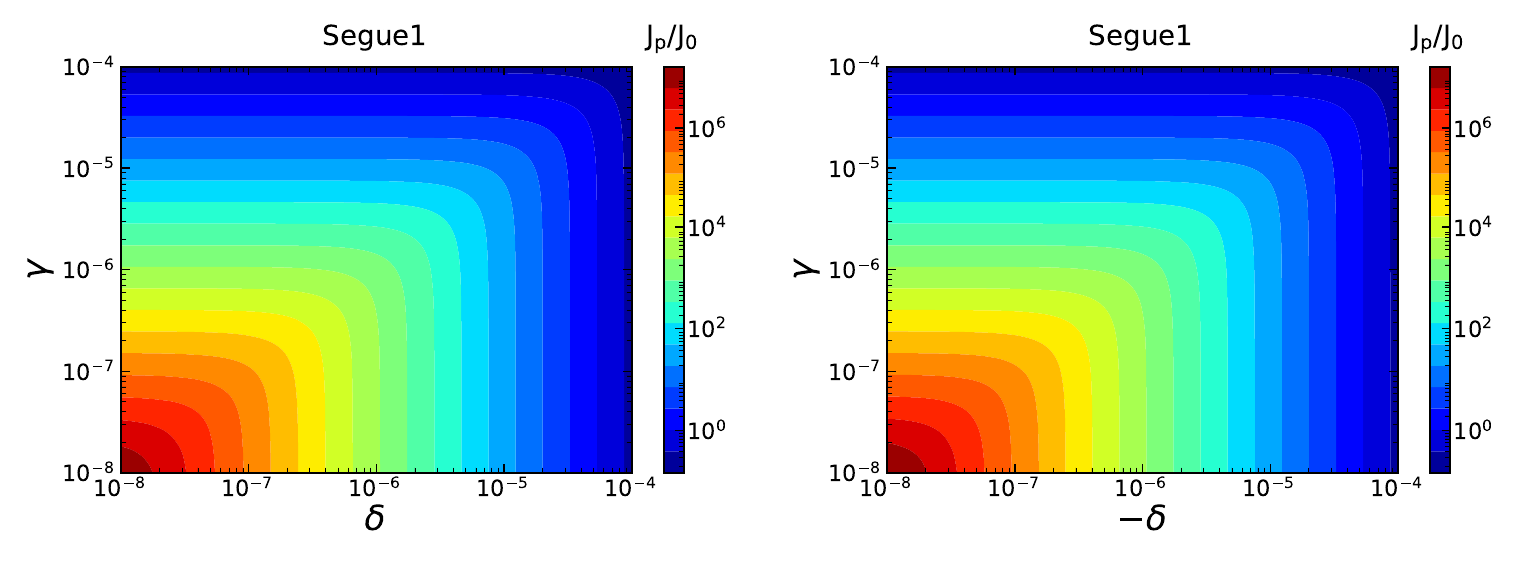}
\caption{The ratio of the $p$-wave Breit-Wigner enhanced $J$-factor to the normal one $J_p/J_0$ of dSph Segue 1 as a function of $\delta$ and $\gamma$. The left panel is for the case of $\delta>0$, and the right panel is for the case of $\delta<0$.
}  
\label{fig:Js}
\end{figure}

We calculate the spherically-symmetric and isotropic phase-space distribution $f_\chi\left(r, v\right)$ of DM particles in 15 nearby dSphs (listed in Tab.~\ref{tab:js}).
The typical velocity of DM particles in these 15 dSphs is $v_{\rm{dSph}}\sim 10-20\ \rm{km\ s^{-1}}$.
The method is based on the Eddington's formula~\cite{2008gady.book.....B} and the assumption of Navarro-Frenk-White (NFW) profile~\cite{Navarro:1996gj} of DM density in dSphs.
The details of the method can be found in our previous work~\cite{Ding:2021zzg}.

Instead of an integral over the line of sight,
the expression of $J_p$ can be rewritten in an integral over the radial distance from the center of the target dSph as
\begin{equation}
\begin{aligned} J_p &=\frac{32 \pi^{3}}{D^{2}} \int_{0}^{r_{\max}} r^{2} d r 
\int_{0}^{v_{\rm{esc}}(r)} v_{1}^{2} f_\chi\left(r, v_{1}\right) d v_{1} \int_{0}^{v_{\rm{esc}}(r)} v_{2}^{2} f_\chi\left(r, v_{2}\right) d v_{2} \\ 
& \times \int_{0}^{\pi} \sin \theta \frac{(v_{1}^{2}+v_{2}^{2}-2 v_{1} v_{2} \cos \theta)/c^2 }{(\delta+(v_{1}^{2}+v_{2}^{2}-2 v_{1} v_{2} \cos \theta)/4c^2)^2+\gamma^2} d \theta\ ,\end{aligned}
\label{eq:jp}
\end{equation}
where $D$ is the distance from the solar system to the center of the dSph under consideration, whose value can be taken from Ref.~\cite{Ackermann:2015zua}.
The integral over solid angle is performed over a circular region with $\Delta\Omega\sim 2.4\times 10^{-4}\ \rm{sr}$, namely, angular radius of $\beta = 0.5^{\circ}$, which corresponds to the maximal radius $r_{\rm{max}}=D \cdot \rm{sin}\beta$.
This integral range is large enough to envelop the region where there is significant DM annihilation for the NFW profile~\cite{Ackermann:2013yva}.
The integration over the DM velocity is cut off at the escape velocity $v_{\mathrm{esc}}(r)$ at radius $r$.

In Tab.~\ref{tab:js}, we show the values of $J_p$ at four typical values of $(\delta, \gamma)=(-10^{-8}, 10^{-8}), (10^{-8}, 10^{-8}), (10^{-6}, 10^{-6}), (10^{-4}, 10^{-4})$ and $J_0$ for 15 nearby dSphs.
The dSphs that have similar $J_0$ factors may have different DM velocity distributions and hence quite different values of $J_p$ factor.
For instance, the value of $J_p$ factor at $(\delta, \gamma)=(10^{-4}, 10^{-4})$ of Ursa Minor ($28.3\times 10^{17}~\mathrm{GeV^2\ cm^{-5}}$) is almost twice the value of Bootes I ($15\times 10^{17}~\mathrm{GeV^2\ cm^{-5}}$), although their $J_0$ factors are nearly the same ($6.3\times 10^{18}~\mathrm{GeV^2\ cm^{-5}}$).
Consequently, the relative importance of $J_p$ factors among the dSphs in their contribution to the gamma-ray flux has changed.
For instance, Ursa Minor (Bootes I) becomes more (less) important in $J_p$ than in $J_0$.
However, Segue 1 (Leo II) remains the brightest (faintest) dSph among all the 15 dSphs.
In Fig.~\ref{fig:Js}, we show the ratio of the $p$-wave Breit-Wigner enhanced $J$-factor $J_p$ to the normal one $J_0$ of dSph Segue 1 as a function of $\delta$ and $\gamma$ in both cases of $\delta>0$ and $\delta<0$.
Approximately, $J_p$ scales as $J_0 \cdot (v_{\rm{dSph}}^2/c^2)/((\delta+v_{\rm{dSph}}^2/c^2)^2+\gamma^2)$.
Hence, $J_p$ increases monotonically with decreasing $\delta$ and $\gamma$ in the case of $\delta>0$.
When $\delta<0$, $J_p$ remains increasing monotonically as $\gamma$ decreases while reaching a maximum at $\delta\sim -v_{\rm{dSph}}^2/c^2$.
$J_p$ is smaller than $J_0$ when $|\delta|, \gamma\gtrsim v_{\rm{dSph}}/c$, but is greater than $J_0$ when $|\delta|, \gamma\lesssim v_{\rm{dSph}}/c$, which can also be seen from Tab.~\ref{tab:js}.
When $|\delta|, \gamma\gg v_{\rm{dSph}}^2/c^2$, $J_p$ has almost the same values in both cases of $\delta>0$ and $\delta<0$.

\subsection{Upper limits on the DM annihilation cross section}
\label{sec:Fer_analysis}

\begin{figure}[t]
\includegraphics[width=\textwidth]{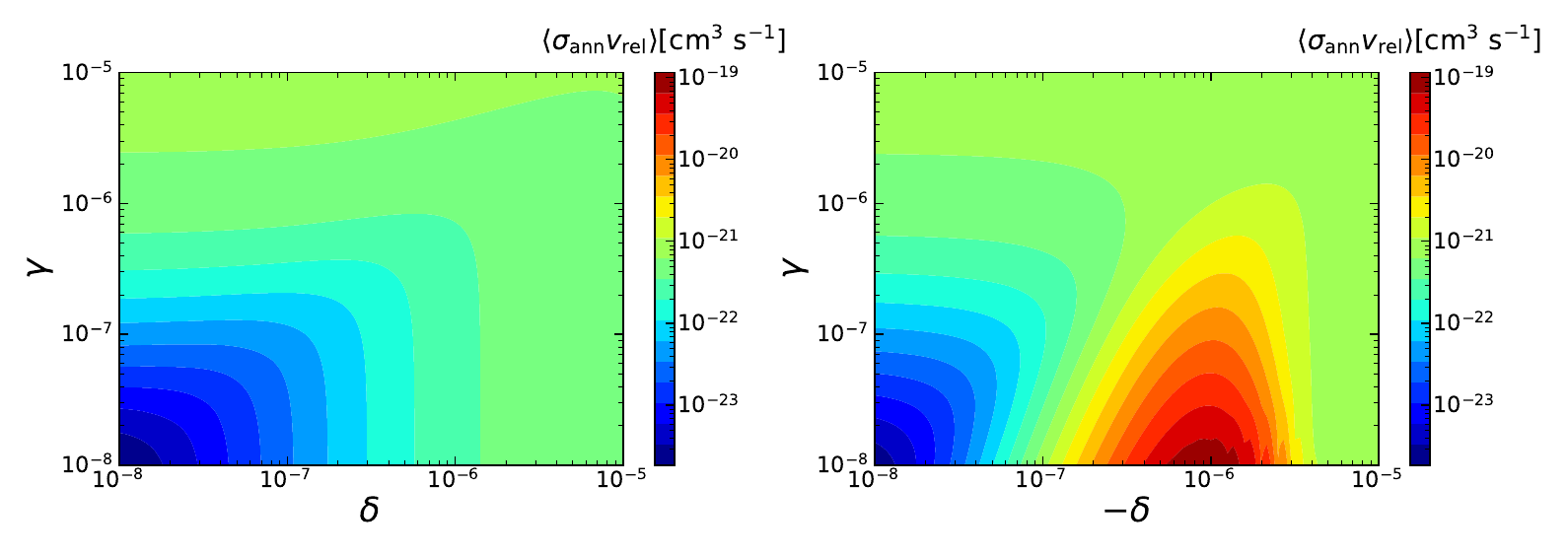}
\caption{
Upper limits on DM annihilation cross section at $95\%$ C.L. as a function of $\delta$ and $\gamma$ derived from a combined analysis on the Fermi-LAT gamma-ray data of the 15 nearby dSphs~\cite{Ackermann:2015zua}. The DM annihilation channel is $\tau^+\tau^-$, and $m_\chi$ is fixed at $1.7\ \rm{TeV}$. The left panel is the for case of $\delta>0$, and the right panel is for the case of $\delta<0$.}
\label{fig:fermi}
\end{figure}

\begin{figure}[t]
\includegraphics[width=\textwidth]{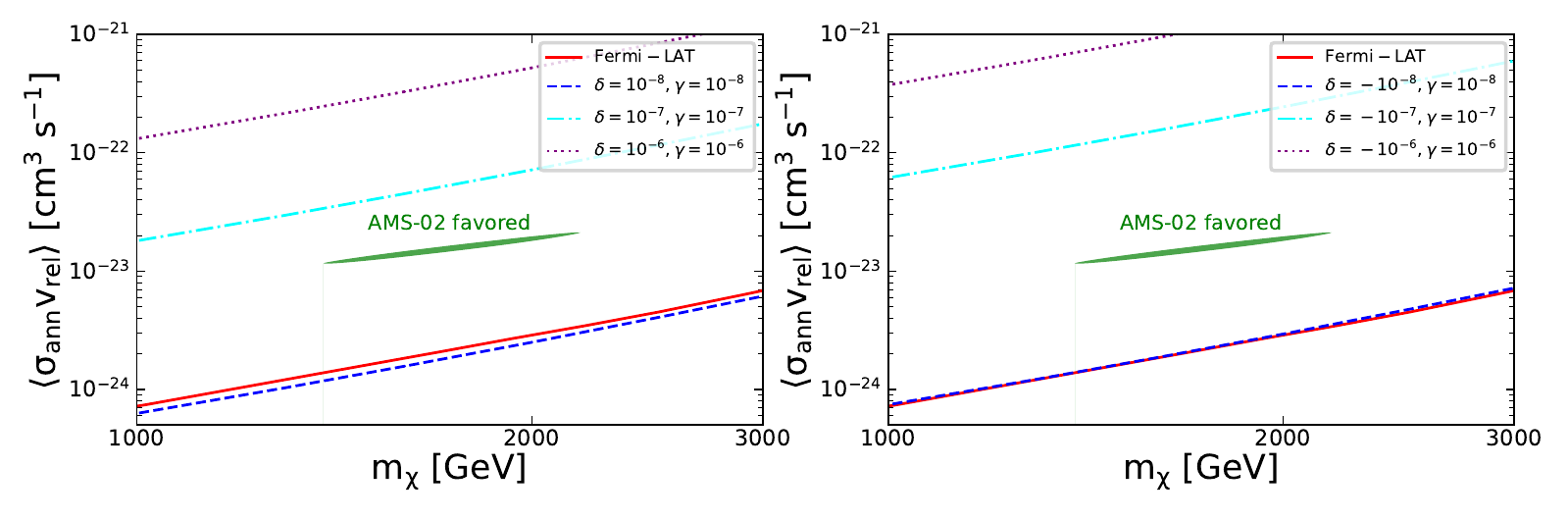}
\caption{
The same upper limits with Fig.~\ref{fig:fermi} as a function of $m_\chi$.
The blue, cyan and purple curves correspond to the parameters of $(|\delta|, \gamma) = (10^{-8}, 10^{-8}), (10^{-7}, 10^{-7})$ and $(10^{-6}, 10^{-6})$, respectively.
The red line is the upper limits on constant DM annihilation cross section, which are taken from Ref.~\cite{Ackermann:2015zua}. 
The left panel is for the case of $\delta>0$, and the right panel is $\delta<0$.
In both panels, the green regions are favored by the AMS-02 CR positron data~\cite{Aguilar:2019owu} at $95\%$ C.L.
}
\label{fig:fer_ams}
\end{figure}

We use the six years of Fermi-LAT PASS8 data on gamma rays from the 15 nearby dSphs listed in Tab.~\ref{tab:js}, and adopt the same approach of standard delta-log-likelihood as in our recent work~\cite{Ding:2021zzg} to place constraints on the velocity-independent part of the DM annihilation cross section $C$.
With the $p$-wave Breit-Wigner enhanced $J$-factor $J_p$ calculated in the previous subsection, it is straightforward to calculate the gamma-ray flux produced from the DM annihilation in dSphs through Eq.~(\ref{eq:gamma_flux}).
We derive the upper limits on $b$ at $95\%$ C.L. from the dSphs gamma-ray data from Fermi-LAT~\cite{Ackermann:2015zua} as a function of $m_\chi$ for specific values of $\delta$ and $\gamma$.
To facilitate the comparison with the AMS-02 data, we convert the obtained upper limits on the value of $b$ to that on $\langle \sigma_{\rm{ann}} v_{\rm{rel}}\rangle$ through Eq.~(\ref{eq:cs}) with $v_0=v_{\rm{halo}}$.

In Fig.~\ref{fig:fermi}, we show the upper limits as a function of $\delta$ and $\gamma$ with DM mass fixed at the best-fit value $m_\chi=1.7\ \rm{TeV}$ for AMS-02 positron data in the $\tau^+\tau^-$ channel in both cases of $\delta>0$ and $\delta<0$.
When $\delta>0$, the upper limits are more stringent at smaller values of $\delta$ and $\gamma$, which can be expected from the adverse behavior of $J_p$ factor.
In the case of $\delta<0$, the most stringent upper limits are located at $\delta\sim -v_{\rm{dSph}}^2/c^2$ and $\gamma\lesssim v_{\rm{dSph}}^2/c^2$, where the values of $J_p$ factor are the largest.
The upper limits reach a maximum at $\delta\sim -v_{\rm{halo}}^2/c^2$ and $\gamma\lesssim v_{\rm{halo}}^2/c^2$ due to the large rescale factor in Eq.~(\ref{eq:cs}) in this region.
Fig.~\ref{fig:fermi} shows that, only within $|\delta|, \gamma\lesssim 10^{-7}$, the constraints from dSphs gamma rays would exclude the AMS-02 favored cross section.
In Fig.~\ref{fig:fer_ams}, we show the upper limits as a function of DM mass at three typical values of $(|\delta|, \gamma)=(10^{-8}, 10^{-8}), (10^{-7}, 10^{-7})$ and $(10^{-6}, 10^{-6})$ in both cases of $\delta>0$ and $\delta<0$.
The upper limits on constant DM annihilation cross section are also shown for comparison.
The figure shows that the upper limits on $p$-wave Breit-Wigner enhanced DM annihilation cross section are much weaker than that on constant cross section at $(|\delta|, \gamma)=(10^{-7}, 10^{-7})$ and $(10^{-6}, 10^{-6})$, and can be compatible with the constraints from AMS-02 positron data.
While for lower values of $(|\delta|, \gamma)=(10^{-8}, 10^{-8})$, the contradiction still exists. 
This is in agreement with Fig.~\ref{fig:fermi}.

\section{Constraints from the CMB}\label{sec:cmb}

\begin{figure}[t]
\includegraphics[width=0.505\textwidth]{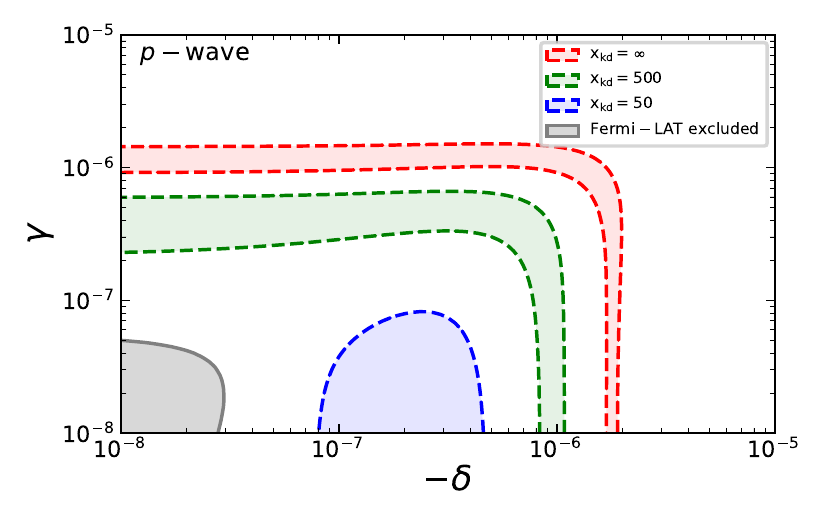}
\includegraphics[width=0.475\textwidth]{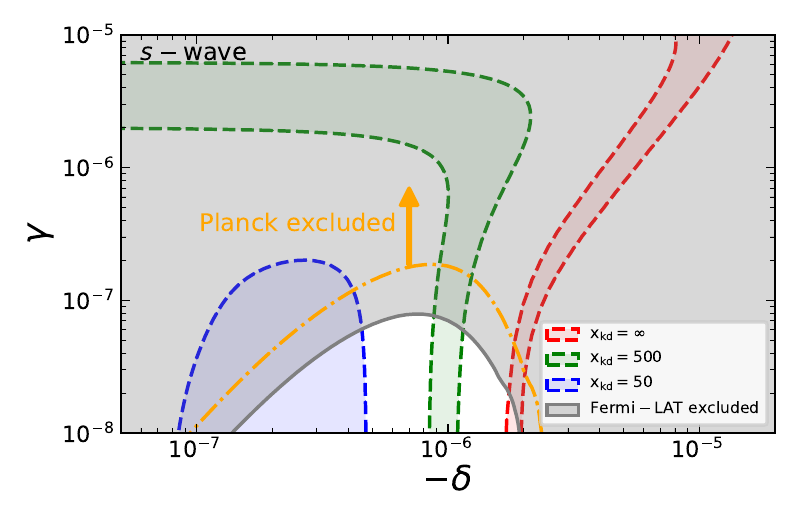}
\caption{
Left) Parameter regions which can simultaneously account for the AMS-02 CR positron excess~\cite{Aguilar:2019owu} and DM thermal relic density~\cite{Aghanim:2018eyx} in the $(\delta, \gamma)$ plane within the scenario of $p$-wave Breit-Wigner enhanced DM annihilation for $\delta<0$.
The blue, green, and red regions correspond to $x_{\rm{kd}}=50,\ 500$ and $\infty$, respectively.
The gray region is excluded by the combined analysis of the Fermi-LAT gamma-ray data of 15 nearby dSphs~\cite{Ackermann:2015zua} at $95\%$~C.L.
The DM annihilation channel is $\tau^+\tau^-$.
Right) The same as the left panel but within the scenario of $s$-wave Breit-Wigner enhanced DM annihilation.
Parameter region above the orange line is excluded by the CMB data from Planck~\cite{Aghanim:2018eyx} at $95\%$~C.L.
}
\label{fig:par_space_neg}
\end{figure}

\begin{figure}[t]
\includegraphics[width=0.7\textwidth]{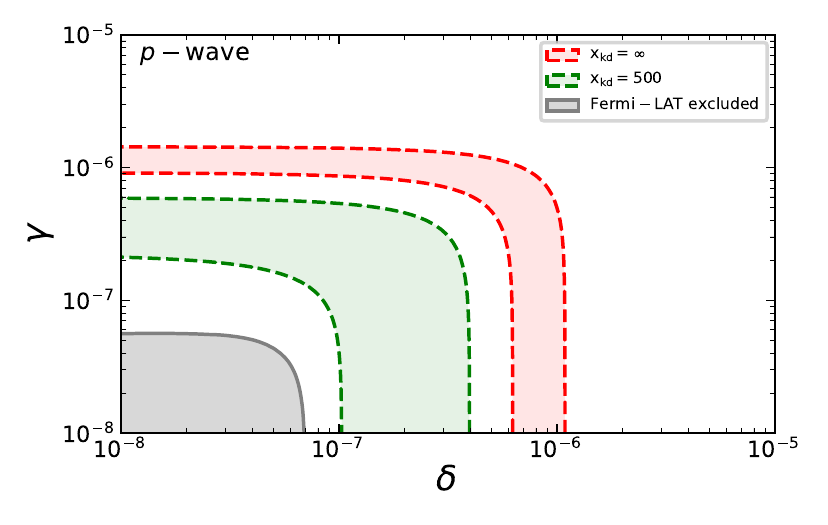}
\caption{
Parameter regions which can simultaneously account for the AMS-02 CR positron excess~\cite{Aguilar:2019owu} and DM thermal relic density~\cite{Aghanim:2018eyx} in the $(\delta, \gamma)$ plane within the scenario of $p$-wave Breit-Wigner enhanced DM annihilation for $\delta>0$.
The green and red regions correspond to $x_{\rm{kd}}=500$ and $\infty$, respectively.
The gray region is excluded by the combined analysis of the Fermi-LAT gamma-ray data of 15 nearby dSphs~\cite{Ackermann:2015zua} at $95\%$~C.L.
The DM annihilation channel is $\tau^+\tau^-$.
}
\label{fig:par_space_pos}
\end{figure}

The recombination history of the Universe can be potentially modified by energy injection into gas, photon-baryon plasma and background radiation from DM annihilation, which can lead to modifications in the temperature and polarization power spectra of CMB~\cite{Chen:2003gz,Padmanabhan:2005es}.
Thus, the measurement on the anisotropy of CMB can be used to constrain the particle nature of DM~\cite{Galli:2009zc,Slatyer:2009yq,Finkbeiner:2011dx}.
An advantage of this indirect detection method over the CR observations is that it is free of some astrophysical uncertainties such as the DM density profile, large-scale structure formation, etc.~\cite{Slatyer:2015jla}.
Using a parameter $f_{\rm{eff}}$ to describe the fraction of the energy released by the DM annihilation process that is transferred to the intergalactic medium around the redshift to which the CMB anisotropy data are most sensitive, namely, $z\sim 600$, the upper limits on the DM annihilation cross section at $95\%$ C.L. recently reported by the Planck collaboration is~\cite{Aghanim:2018eyx}
\begin{equation}\label{eq:planck}
f_{\rm{eff}}\langle \sigma_{\rm{ann}} v_{\rm{rel}} \rangle/m_{\chi}\leq 3.2 \times 10^{-28}\ \mathrm{cm^3\ s^{-1}\ GeV^{-1}}\ .
\end{equation}
We take the value of $f_{\rm{eff}}\approx 0.16$ for $\mu^+\mu^-$ and $\approx 0.15$ for $\tau^+\tau^-$ channel at $m_\chi\sim \mathcal{O}(1)\ \rm{TeV}$~\cite{Madhavacheril:2013cna}.
Assuming a velocity-independent DM annihilation cross section, we show in Fig.~\ref{fig:ams-pos} the upper limits at $95\%$ C.L. on DM annihilation cross section as a function of DM mass from the CMB data, which indicate strong contradiction with the AMS-02 favored parameter regions.
For the $p$-wave Breit-Wigner enhanced DM annihilation, the cross section is expected to be strongly suppressed by the velocity-square factor due to the extremely low DM velocity of $\sim 1\ \mathrm{m\cdot s^{-1}}$ in the epoch of recombination, which results in rather weak constraints from CMB.

\section{Combined analysis}\label{sec:result}

\begin{figure}[t]
\includegraphics[width=0.7\textwidth]{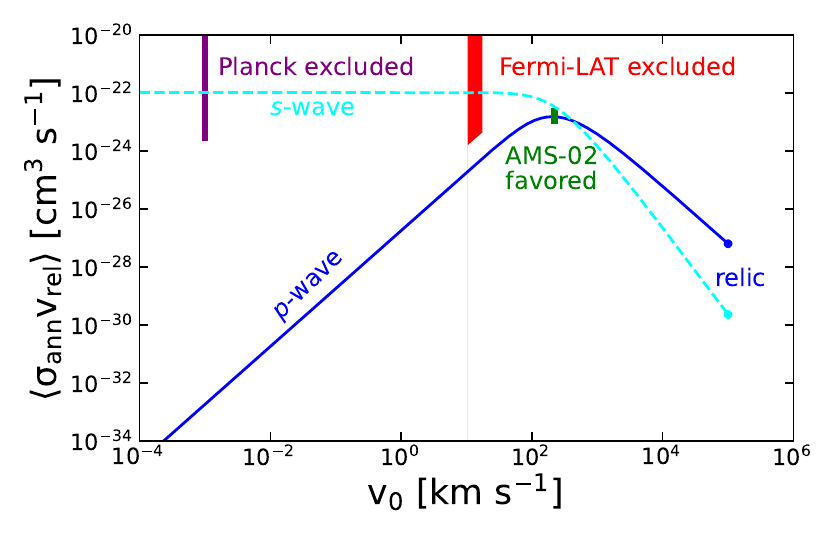}
\caption{
$p$-wave Breit-Wigner enhanced DM annihilation cross section as a function of the DM most probable velocity $v_0$ with the global factor $b$ fixed by the DM thermal relic density (blue solid). The DM annihilation channel is $\tau^+\tau^-$, and the parameters are set at $\delta=2\times 10^{-7},\ \gamma=2\times 10^{-7},\ m_\chi=1.7\ \rm{TeV}$, and $x_{\rm{kd}}=500$.
For comparison, the case of $s$-wave Breit-Wigner enhanced DM annihilation is also shown (cyan dashed).
The green region represents the cross section favored by AMS-02 positron data~\cite{Aguilar:2019owu} at $v_0\sim 220\ \mathrm{km\ s^{-1}}$.
The red region corresponds to the upper limits at $95\%$~C.L. from a combined analysis of the Fermi-LAT gamma-ray data from the 15 nearby dSphs~\cite{Ackermann:2015zua} with $v_0\sim 10-20\ \mathrm{km\ s^{-1}}$.
The purple vertical line indicates the cross section excluded by Planck CMB observations~\cite{Aghanim:2018eyx} at $95\%$ C.L. with $v_0\sim 1\ \mathrm{m\ s^{-1}}$.
}
\label{fig:csv0}
\end{figure}

To identify the parameter space that can simultaneously explain the AMS-02 positron excess and reproduce the correct DM thermal relic density while evading the constraints from dSphs gamma rays and CMB in the $p$-wave Breit-Wigner enhanced DM annihilation, we perform a scan in the $(\delta, \gamma)$ plane.
We consider three typical DM kinetic decoupling temperatures of $x_{\rm{kd}}=50, 500, \infty$ as benchmark values.
In the left panel of Fig.~\ref{fig:par_space_neg}, we show the constraints at $95\%$ C.L. from these observations on parameter regions for the case of $\delta<0$.
To give the required boost factor which can account for the CR positron excess, the allowed parameters are limited in a narrow range within $|\delta|, \gamma \lesssim 10^{-6}$, which depends on $x_{\rm{kd}}$.
A lower kinetic decoupling temperature allows larger values of $|\delta|$ and $\gamma$.
The parameter regions within $|\delta|, \gamma \lesssim 10^{-7}$ are excluded by the constraints from dSphs gamma rays.
Due to the strong velocity suppression of the annihilation cross section during recombination, the CMB data imposes almost no additional constraints to the parameter regions.
As a comparison, we also show the results of $s$-wave Breit-Wigner enhanced DM annihilation in the right panel of Fig.~\ref{fig:par_space_neg}, which perform quite differently.
It shows that most of the parameter regions that can provide the required boost factor are excluded by the CMB data, and dSphs gamma-ray data places stricter constraints than CMB.
Due to the monotonic increase in the DM annihilation cross section with decreasing DM velocity in the case of $\delta>0$, the $s$-wave Breit-Wigner enhancement cannot accommodate the observations of CR positron excess, dSphs gamma rays, and CMB simultaneously.
While in $p$-wave Breit-Wigner enhanced DM annihilation, $\delta>0$ is still allowable.
The corresponding allowed parameter regions are shown in Fig.~\ref{fig:par_space_pos}, which are similar with the case of $\delta<0$ in the left panel of Fig.~\ref{fig:par_space_neg} except that the Breit-Wigner enhancement is not large enough to explain the AMS-02 positron excess when the DM kinetic decoupling temperature has the same scale as the freeze-out temperature ($x_{\rm{kd}}=50$).

According to the results above, only tiny values of $|\delta|$ and $\gamma$ ($\lesssim 10^{-6}$) can survive all the constraints mentioned above.
The narrow decay width of the resonance can be realized through weak coupling of the resonance with other particles.
As for how to naturally realize the tiny deviation of the resonance mass from the threshold,
one possible way proposed in Ref.~\cite{Bai:2017fav} is through the nontrivial flavour symmetry-breaking in the dark sector, where a resonance with a mass of almost $2m_\chi$ is realized by assigning a particular symmetry-breaking mode and the tiny mass deviation is induced by loop effect.

To illustrate how the DM annihilation cross section varies in different epochs and regions of the Universe, in Fig.~\ref{fig:csv0}, we show $\langle \sigma_{\rm{ann}} v_{\rm{rel}} \rangle$ as a function of the velocity $v_0$ through Eq.~(\ref{eq:cs}) for a typical parameter set of $m_\chi=1.7\ \mathrm{TeV}$, $\delta=2\times 10^{-7}$, $\gamma=2\times 10^{-7}$, and $x_{\rm{kd}}=500$.
To reproduce the correct DM thermal relic density, the cross section at freeze-out ($v_0\sim c/3$) is fixed at $\sim 6.3\times 10^{-28}~\mathrm{cm^3\ s^{-1}}$.
When the velocity decreases, the cross section increases as $c^2/v_0^2$ due to the Breit-Wigner enhancement, and reaches a maximum of $\sim 1.5\times 10^{-23}~\mathrm{cm^3\ s^{-1}}$ at $v_0\sim 220~\mathrm{km\cdot s^{-1}}$, which is large enough to provide the AMS-02 positron excess.
As the velocity decreases further, the Breit-Wigner enhancement saturates and the cross section decreases as $v_0^2/c^2$.
The cross section at $v_0\sim 10~\mathrm{km\cdot s^{-1}}$ is almost one order of magnitude smaller than the upper limits derived from a combined analysis of Fermi-LAT gamma-ray data on 15 nearby dSphs.
Finally, the cross section at $v_0\sim 1~\mathrm{m\cdot s^{-1}}$ becomes so low that it no longer suffers from the constraints from CMB observations.
We also show the velocity dependence of the $s$-wave Breit-Wigner enhanced DM annihilation cross section with the same parameter set, whose value at freeze-out is also fixed by the observed DM thermal relic density.
The cross section increases as $c^4/v_0^4$ and then remains almost constant as the velocity decreases.
The result shows that although the $s$-wave Breit-Wigner enhancement can account for the large boost factor in the case of $\delta>0$, it is incompatible with the observations of dSphs gamma rays and CMB due to the monotonic increase in annihilation cross section with decreasing DM velocity.

\section{Conclusions}\label{sec:conclusion}

In summary, $p$-wave Breit-Wigner enhanced DM annihilation can have complicated velocity dependence compared with that for the $s$-wave case.
We have taken into account the impact of the kinetic decoupling effect on DM thermal relic density.
We have systematically calculated the velocity-dependent $J$-factor for the $p$-wave Breit-Wigner enhanced DM annihilation for 15 nearby dSphs and derived the corresponding upper limits of the DM annihilation cross section through a combined analysis of the dSphs gamma-ray data.
We have identified the parameter regions that can simultaneously explain the AMS-02 positron excess and the DM thermal relic density while evading the constraints from dSphs gamma rays and CMB.
Our results show that the allowed parameter regions in $p$-wave Breit-Wigner enhanced DM annihilation are larger than that in the case of $s$-wave.
In addition, whether the resonance mass is above or below the threshold, there exist parameter regions that can explain these observations consistently.

\acknowledgements
This work is supported in part by
The National Key R\&D Program of China No.~2017YFA0402204,
the Key Research Program of the Chinese Academy of Sciences (CAS), Grant NO. XDPB15,
the CAS Project for Young Scientists in Basic Research YSBR-006, and
the National Natural Science Foundation of China (NSFC)
No.~11825506,  No.~11821505, and  No.~12047503.  

\bibliographystyle{arxivref}
\bibliography{PwaveBreitWigner}

\end{document}